\documentclass[prd,aps,floats,twocolumn]{revtex4}

\usepackage[dvips]{graphicx}
\usepackage{amssymb}
\usepackage{amsmath}

\usepackage{color}

\newcommand{\TE}{\Tilde{E}}
\newcommand{\TL}{\Tilde{L}}
\newcommand{\Tr}{\Tilde{r}}
\newcommand{\Ta}{\Tilde{a}}
\newcommand{\Ttau}{\Tilde{\tau}}
\newcommand{\Tt}{\Tilde{t}}
\newcommand{\Tx}{\Tilde{x}}
\newcommand{\TO}{\Tilde{\Omega}}
\newcommand{\Tp}{T_{\rm plunge}}
\newcommand{\RC}{\dot{\mathcal{E}}}

\begin{document}

\title{Transition from adiabatic inspiral to plunge into a spinning black hole}

\author{Michael Kesden} \email{mhk10@nyu.edu}

\affiliation{Center for Cosmology and Particle Physics, Department of Physics,
New York University, 4 Washington Pl., New York, New York 10003, USA}

\date{January 2011}

\begin{abstract}
A test particle of mass $\mu$ on a bound geodesic of a Kerr black hole
of mass $M \gg \mu$ will slowly inspiral as gravitational radiation
extracts energy and angular momentum from its orbit.  This inspiral
can be considered {\it adiabatic} when the orbital period is much
shorter than the timescale on which energy is radiated, and {\it
quasicircular} when the radial velocity is much less than the
azimuthal velocity.  Although the inspiral always remains adiabatic
provided $\mu \ll M$, the quasicircular approximation breaks down as
the particle approaches the innermost stable circular orbit (ISCO).
In this paper, we relax the quasicircular approximation and solve the
radial equation of motion explicitly near the ISCO.  We use the
requirement that the test particle's 4-velocity remains properly
normalized to calculate a new contribution to the difference between
its energy and angular momentum.  This difference determines how a
black hole's spin changes following a test-particle merger, and can be
extrapolated to help predict the mass and spin of the final black hole
produced in finite-mass-ratio black-hole mergers.  Our new
contribution is particularly important for nearly maximally spinning
black holes, as it can affect whether a merger produces a naked
singularity.
\end{abstract}

\maketitle

\section{Introduction} \label{S:intro}

Supermassive black holes (SBHs) with masses $10^6 M_\odot \lesssim M
\lesssim 10^{10} M_\odot$ reside at the centers of most large galaxies
\cite{Kormendy:1995er}.  These SBHs will be surrounded by dense cusps
of stars \cite{Bahcall:1976aa}, some fraction of which will consist of
compact objects (white dwarfs, neutron stars, or stellar-mass black
holes) of mass $\mu \sim 1 - 10~M_\odot$.  Compact objects whose
orbital velocities lie within a ``loss cone'' about the radial
direction \cite{Frank:1976uy} will inspiral into the SBHs under the
influence of gravitational radiation \cite{Sigurdsson:1996uz}.  These
extreme-mass-ratio inspirals (EMRIs) will be an important source of
gravitational waves (GWs) for the proposed space-based GW detector
LISA \cite{LISA}.  LISA will rely on matched filtering to detect
EMRIs, requiring templates that accurately track the phase of the GWs
when their fundamental frequency $f$ is in the range $10^{-4} \lesssim
f \lesssim 10^{-1}$ Hz.  These GW frequencies correspond to orbital
frequencies $\Omega = \pi f$ at the innermost stable circular orbits
(ISCOs) of SBHs with masses $10^5 M_\odot \lesssim M \lesssim 10^7
M_\odot$.  To maximize the number of EMRIs detected by LISA, it is
therefore essential to understand the trajectories of compact objects
as they inspiral all the way to the ISCO.

In the absence of gravitational radiation, test particles travel on
geodesics of the Kerr metric $g_{\mu\nu}$ \cite{Kerr:1963ud} that
describes the spacetime of spinning black holes.  Kerr geodesics are
characterized by four constants of motion: the particle's rest mass
$\mu$, energy $E$, $z$ component of angular momentum $L_z$, and Carter
constant $Q$ \cite{Carter:1968rr}.  These constants can be determined
from the particle's 4-momentum $p^\mu$ and the Kerr metric's timelike
Killing vector $T_\mu$, azimuthal Killing vector $\Phi_\mu$, and
Killing tensor $Q_{\mu\nu}$ \cite{Walker:1970un}:
\begin{subequations} \label{E:4con}
  \begin{eqnarray} \label{E:masscon}
	\mu &=& \sqrt{-g_{\mu\nu} p^\mu p^\nu}~, \\ \label{E:Econ}
	E &=& -T_\mu p^\mu~, \\ \label{E:Lcon}
	L_z &=& \Phi_\mu p^\mu~, \\ \label{E:Qcon}
	Q &=& Q_{\mu\nu} p^\mu p^\nu~.
  \end{eqnarray}
\end{subequations}
Equatorial geodesics have $Q = 0$; for computational simplicity we
will restrict our attention to equatorial orbits for the remainder of
this paper.

In the test-particle limit $\eta \equiv \mu/M \ll 1$, the
stress-energy tensor of a test particle moving on a Kerr geodesic
sources gravitational radiation that can be calculated using
black-hole perturbation theory \cite{Teukolsky:1973ha}.  This
radiation extracts energy and angular momentum from the orbit, causing
the particle to migrate through the phase space $\{ E, L_z \}$.  The
GW energy flux $\dot{E}_{\rm GW}$ and angular momentum flux
$\dot{L_z}_{\rm GW}$ (an overdot symbolizes a derivative with respect
to Boyer-Lindquist coordinate time) are proportional to $\eta^2$,
implying that the timescale $t_{\rm GW} \sim E/\dot{E}_{\rm GW}$ is
proportional to $\eta^{-1}$.  The orbital period $t_{\rm orb} \sim
\Omega^{-1}$ is independent of $\eta$ to lowest order, implying that
the inspiral will always be {\it adiabatic} ($t_{\rm GW} \gg t_{\rm
orb}$) for sufficiently small $\eta$.  Gravitational radiation
circularizes eccentric orbits in the post-Newtonian regime
\cite{Peters:1963ux}, implying that the early inspiral will also be
{\it quasicircular} ($\dot{r} \ll r\Omega$).  During this adiabatic,
quasicircular portion of the inspiral, the trajectory of the test
particle is well described by a sequence of circular geodesics of ever
decreasing Boyer-Lindquist radius $r$.  In the quasicircular
approximation, the radial velocity is given by
\begin{equation} \label{E:adRV}
\dot{r} = -\frac{\dot{E}_{\rm GW}}{dE/dr}~,
\end{equation}
where $dE/dr$ is the derivative of the energy of a circular equatorial
geodesic with respect to its Boyer-Lindquist radius.  This radial
velocity is proportional to $\eta$, and approaches negative infinity
as $r$ approaches $r_{\rm ISCO}$ where $dE/dr$ vanishes.  This
behavior is unphysical, and reflects the breakdown of the
quasicircular approximation as the test particle approaches the ISCO.

In the vicinity of the ISCO, the test particle transitions from the
quasicircular inspiral described above to a ``captured'' plunge that
crosses the event horizon \cite{Bardeen:1972fi}.  This {\it transition
regime} has been investigated in the test-particle limit by Ori and
Thorne \cite{Ori:2000zn} (hereafter OT) and for nonspinning black
holes or arbitrary mass ratio by Buonanno and Damour
\cite{Buonanno:2000ef}.  Sundararajan \cite{Sundararajan:2008bw}
generalized the approach of OT to inclined and eccentric orbits,
deriving trajectories that served as sources for the gravitational
waveforms calculated in Sundararajan, Khanna, and Hughes
\cite{Sundararajan:2010sr}.  We will adopt the notation of OT
throughout this paper to facilitate comparisons between our results
and theirs.  In Sec.~\ref{S:prev}, we will review the previous
treatment of the transition regime and the discovery of scaling
relations that can be used to apply a universal dimensionless
trajectory to mergers with arbitrary mass ratio and black-hole spin.
In Sec.~\ref{S:4v}, we show that the test particle's 4-momentum
$p^\mu$ is not properly normalized according to Eq.~(\ref{E:masscon})
in this model of the transition.  We develop a new model that
satisfies this requirement by allowing the particle's energy and
angular momentum to vary independently.  We then explore this new
model's predictions for the behavior of the transition regime for
nearly maximal black-hole spins in Sec.~\ref{S:maxspins}.  A summary
of our principle findings and their implications is given in
Sec.~\ref{S:disc}.

\section{Previous Treatment} \label{S:prev}

\subsection{Geodesic Motion} \label{SS:geo}

We begin by reviewing the motion of test particles on Kerr geodesics.
In the equatorial plane ($\theta = \pi/2$), the Kerr metric for a
black hole of mass $M$ and spin $a$ can be written in Boyer-Lindquist
coordinates \cite{Boyer:1966qh} as
\begin{multline} \label{E:KerrEq}
ds^2 = -\left( 1 - \frac{2M}{r} \right) dt^2
+ \left( 1 - \frac{2M}{r} + \frac{a^2}{r^2} \right)^{-1} dr^2 \\
+ r^2 d\theta^2
+ \left (r^2 + a^2 + \frac{2Ma^2}{r} \right) d\phi^2
- \frac{4Ma}{r} dt d\phi~.
\end{multline}
Here and throughout this paper we use units in which Newton's constant
$G$ and the speed of light $c$ are unity.  We can also define a
dimensionless radius $\Tr \equiv r/M$, coordinate time $\Tt \equiv
t/M$, and spin $\Ta \equiv a/M$.  In these coordinates, the three
constants of motion given in Eqs.~(\ref{E:masscon}) through
(\ref{E:Lcon}) provide three equations for the evolution of $\Tt$,
$\Tr$, and $\phi$ as functions of the dimensionless proper time $\Ttau
\equiv \tau/M$ along the particle's worldline.  Eq.~(\ref{E:masscon})
can be rewritten as
\begin{eqnarray} \label{E:4norm}
1 &=& -g_{\mu\nu} \frac{d\Tx^\mu}{d\Ttau} \frac{d\Tx^\nu}{d\Ttau}
\nonumber \\
&=& \left( 1 - \frac{2}{\Tr} \right) \left( \frac{d\Tt}{d\Ttau} \right)^2
- \left( 1 - \frac{2}{\Tr} + \frac{\Ta^2}{\Tr^2} \right)^{-1}
\left( \frac{d\Tr}{d\Ttau} \right)^2 \nonumber \\
&& - \left (\Tr^2 + \Ta^2 + \frac{2\Ta^2}{\Tr} \right)
\left( \frac{d\phi}{d\Ttau} \right)^2 + \frac{4\Ta}{\Tr}
\frac{d\Tt}{d\Ttau} \frac{d\phi}{d\Ttau}~,
\end{eqnarray}
while according to Eqs.~(\ref{E:Econ}) and (\ref{E:Lcon}), the
dimensionless energy $\TE \equiv E/\mu$, and angular momentum $\TL
\equiv L_z/\mu M$ are given by
\begin{subequations} \label{E:DEL}
  \begin{eqnarray} \label{E:DE}
	\TE &=& \left( 1 - \frac{2}{\Tr} \right) \frac{d\Tt}{d\Ttau}
	+ \frac{2\Ta}{\Tr} \frac{d\phi}{d\Ttau}~, \\ \label{E:DL}
	\TL &=& \left (\Tr^2 + \Ta^2 + \frac{2\Ta^2}{\Tr} \right)
	\frac{d\phi}{d\Ttau} - \frac{2\Ta}{\Tr} \frac{d\Tt}{d\Ttau}~. 
  \end{eqnarray}
\end{subequations}
Solving Eqs.~(\ref{E:DE}) and (\ref{E:DL}) for $d\Tt/d\Ttau$ and
$d\phi/d\Ttau$, inserting the result into Eq.~(\ref{E:4norm}), then
solving for $(d\Tr/d\Ttau)^2$ yields
\begin{equation} \label{E:RV2}
\left( \frac{d\Tr}{d\Ttau} \right)^2 = \TE^2 - V(\Tr, \TE, \TL)~,
\end{equation}
where
\begin{equation} \label{E:pot}
V(\Tr, \TE, \TL) \equiv 1 - \frac{2}{\Tr}
+ \frac{\TL^2 + \Ta^2 - \TE^2 \Ta^2}{\Tr^2} - \frac{2(\TL - \TE \Ta)^2}{\Tr^3}
\end{equation}
is the effective potential.

Geodesic motion is alternatively described by the geodesic equations
\begin{equation} \label{E:GE}
\frac{d^2\Tx^\mu}{d\Ttau^2} + \Gamma^{\mu}_{\alpha\beta} 
\frac{d\Tx^\alpha}{d\Ttau} \frac{d\Tx^\beta}{d\Ttau} = 0~,
\end{equation}
where $\Gamma^{\mu}_{\alpha\beta}$ are the Christoffel symbols for the
Kerr metric.  When $\TE$ and $\TL$ are constants, the second of these
equations ($\Tx^1 = \Tr$) is equivalent to the derivative of
Eq.~(\ref{E:RV2}) with respect to $\Ttau$
\begin{equation} \label{E:RA}
\frac{d^2\Tr}{d\Ttau^2} = -\frac{1}{2} \frac{\partial V}{\partial \Tr}~.
\end{equation}

\subsection{Near the ISCO} \label{SS:nIsco}

In the vicinity of the ISCO, Eq.~(\ref{E:RV2}) can be Taylor expanded
about the ISCO values of $\Tr$, $\TE$, and $\TL$.  Expanding in the
small variables
\begin{subequations} \label{E:pert}
  \begin{eqnarray} \label{E:rpert}
	R &\equiv& \Tr - \Tr_{\rm ISCO}~, \\ \label{E:Epert}
	\chi &\equiv& \TO^{-1}_{\rm ISCO} (\TE - \TE_{\rm ISCO})~,
	\\ \label{E:Lpert}
	\xi &\equiv& \TL - \TL_{\rm ISCO}~,
  \end{eqnarray}
\end{subequations}
where $\TO_{\rm ISCO}$ is the orbital frequency at the ISCO, yields
\begin{eqnarray} \label{E:1OT}
	\left( \frac{dR}{d\Ttau} \right)^2 &=& - \frac{2\alpha}{3} R^3
	+ 2\beta R \xi + \frac{\partial V}{\partial \TL} (\chi - \xi)
	\nonumber \\
	&& - \TO \frac{\partial^2 V}{\partial \TE \partial \Tr} (\chi - \xi)R
	+~...
\end{eqnarray}
where the ellipsis denotes higher-order terms.  Following the OT
notation we define
\begin{subequations} \label{E:OTco}
  \begin{eqnarray} \label{E:OTalpha}
	\alpha &\equiv& \frac{1}{4} \left(
	\frac{\partial^3 V}{\partial \Tr^3} \right)_{\rm ISCO}~,
	\\ \label{E:OTbeta}
	\beta &\equiv& -\frac{1}{2} \left(
	\frac{\partial^2 V}{\partial \TL \partial \Tr} +
	\TO \frac{\partial^2 V}{\partial \TE \partial \Tr}
	\right)_{\rm ISCO}~,
  \end{eqnarray}
\end{subequations}
and we have made use of the relation
\begin{equation} \label{E:con1}
\TE - \frac{1}{2} \frac{\partial V}{\partial \TE} - \frac{1}{2}
\TO^{-1} \frac{\partial V}{\partial \TL} = 0
\end{equation}
which holds at extrema of the effective potential.  Again assuming
that $\TE$ and $\TL$ are constant, Eq.~(\ref{E:1OT}) is equivalent to
\begin{equation} \label{E:2OT}
\frac{d^2R}{d\Ttau^2} = - \alpha R^2 + \beta \xi
- \frac{1}{2} \TO \frac{\partial^2 V}{\partial \TE \partial \Tr}
(\chi - \xi)~.
\end{equation}
Terms of $\mathcal{O}(R^3)$, $\mathcal{O}(\xi^2, \chi^2)$ and higher
order have been neglected in Eq.~(\ref{E:2OT}).

\subsection{Radiation Reaction} \label{SS:RR}

We have so far neglected the effects of radiation reaction on the test
particle's motion.  The proper way to include radiation reaction would
be to calculate the conservative and dissipative parts of the
self-force \cite{Mino:1996nk}, and add these additional terms to the
right-hand side of Eq.~(\ref{E:GE}).  Although self-force calculations
have progressed rapidly in recent years (for recent reviews see
\cite{PoissLiv,Barack:2009ux}), the full self-force is not yet available
for test particles on Kerr geodesics near the ISCO.  Without this
self-force, the effects of radiation reaction can be approximated by
making the energy and angular momentum time dependent in
Eqs.~(\ref{E:RV2}) and (\ref{E:RA}).

For particles on circular orbits, the rates at which energy and
angular momentum are radiated are related by
\begin{equation} \label{E:circrad}
\frac{d\TE}{d\Ttau} = \TO \frac{d\TL}{d\Ttau}~.
\end{equation}
OT assume that this relation holds throughout the transition,
implying that
\begin{equation} \label{E:xicon}
\chi = \xi = -\eta \kappa \Ttau~,
\end{equation}
where
\begin{equation} \label{E:kappadef}
\kappa \equiv \left( \TO^{-1} \frac{d\TE}{d\Tt} \frac{d\Tt}{d\Ttau}
\right)_{\rm ISCO}~.
\end{equation}
We will later relax this assumption in Sec.~\ref{S:4v}.  Inserting
Eq.~(\ref{E:xicon}) into Eq.~(\ref{E:2OT}) yields
\begin{equation} \label{E:2OTc}
\frac{d^2R}{d\Ttau^2} = - \alpha R^2 - \eta \beta \kappa \Ttau~,
\end{equation}
identical to Eq.~(3.15) of OT.

\subsection{Dimensionless Equation of Motion} \label{SS:dimless}

\begin{figure}[t!]
\begin{center}
\includegraphics[width=3.5in]{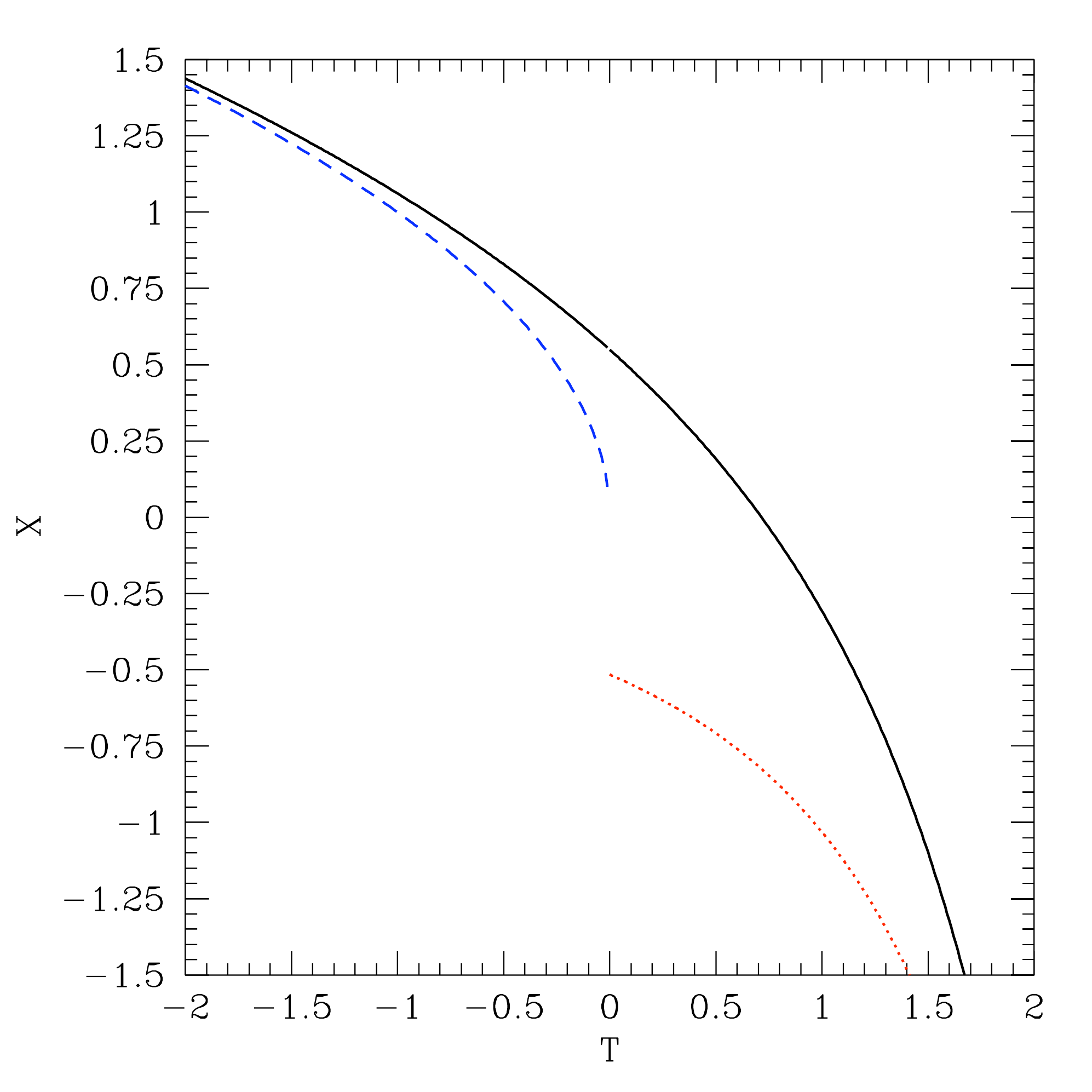}
\end{center}
\caption{The dimensionless radius $X$ as a function of dimensionless
time $T$ during the transition from adiabatic inspiral to plunge.  The
solid black curve shows the numerical solution to
Eq.~(\ref{E:dimEOM}), while the dashed blue and dotted red curves show
the approximate analytic solutions at early and late times given by
Eqs.~(\ref{E:diminsp}) and (\ref{E:dimplunge}) respectively.  This
figure is a reproduction of Fig. 2 in OT.}
\label{F:dimtraj}
\end{figure}

Previous studies \cite{Ori:2000zn,Buonanno:2000ef} noticed that
Eq.~(\ref{E:2OTc}) can be converted into dimensionless form by
defining
\begin{subequations} \label{E:dimvar}
  \begin{eqnarray} \label{E:dimR}
	R &\equiv& \eta^{2/5} R_0 X~, \\ \label{E:dimtau}
	\Ttau &\equiv& \eta^{-1/5} \tau_0 T~,
  \end{eqnarray}
\end{subequations}
where
\begin{subequations} \label{E:scal}
  \begin{eqnarray} \label{E:Rscal}
	R_0 &=& (\beta \kappa)^{2/5} \alpha^{-3/5}~, \\ \label{E:tauscal}
	\tau_0 &=& (\alpha \beta \kappa)^{-1/5}~.
  \end{eqnarray}
\end{subequations}
In these variables, Eq.~(\ref{E:2OTc}) becomes
\begin{equation} \label{E:dimEOM}
\frac{d^2X}{dT^2} = - X^2 - T~.
\end{equation}
At early times, the particle's radial velocity and acceleration
approach zero, suggesting that the correct solution to
Eq.~(\ref{E:dimEOM}) asymptotes to
\begin{equation} \label{E:diminsp}
X = \sqrt{-T} \quad \quad T \to -\infty~.
\end{equation}
At late times, the particle plunges into the horizon ($X = -\infty$)
in a finite proper time $\Ttau$.  The second term on the right-hand of
Eq.~(\ref{E:dimEOM}) can then be neglected, yielding the approximate
solution
\begin{equation} \label{E:dimplunge}
X = \frac{-6}{(\Tp - T)^2} \quad \quad T \to \Tp~.
\end{equation}
Numerically, $\Tp = 3.412$.  The numerical solution to
Eq.~(\ref{E:dimEOM}) and approximate analytic solutions of
Eqs.~(\ref{E:diminsp}) and (\ref{E:dimplunge}) are given by the solid
black, dashed blue, and dotted red curves in Fig.~\ref{F:dimtraj}.

In addition to the approximate radial trajectory $X(T)$, this model
also provides an estimate of the energy and angular momentum radiated
during the transition.  Assuming that energy and angular momentum are
radiated at the ISCO rate throughout the transition,
Eq.~(\ref{E:dimplunge}) implies that the test particle will radiate an
additional amount
\begin{subequations} \label{E:final}
  \begin{eqnarray} \label{E:Efinal}
	\Delta \TE_{\rm tr} &\equiv& \TE_{\rm ISCO} - \TE_{\rm final} = 
	\eta^{4/5} \TO_{\rm ISCO} \kappa \tau_0 \Tp \quad \\ \label{E:Lfinal}
	\Delta \TL_{\rm tr} &\equiv& \TL_{\rm ISCO} - \TL_{\rm final} = 
	\eta^{4/5} \kappa \tau_0 \Tp
  \end{eqnarray}
\end{subequations}
beyond that calculated in the adiabatic approximation.  Since $\Delta
\TE_{\rm tr}$ and $\Delta \TL_{\rm tr}$ scale as $\eta^{4/5}$, in the
test-particle limit $\eta \to 0$ the specific energy $\TE$ and angular
momentum $\TL$ radiated during the transition should exceed that
radiated during the ringdown which scales as $\eta$.  Given the
success of efforts to predict black-hole spins by extrapolating from
the test-particle limit
\cite{Hughes:2002ei,Buonanno:2007sv,Kesden:2008ga,Kesden:2009ds}, a
closer examination of the energy and angular momentum radiated during
the transition is worthwhile.

\section{Normalization of 4-velocity} \label{S:4v}

\begin{figure}[t!]
\begin{center}
\includegraphics[width=3.5in]{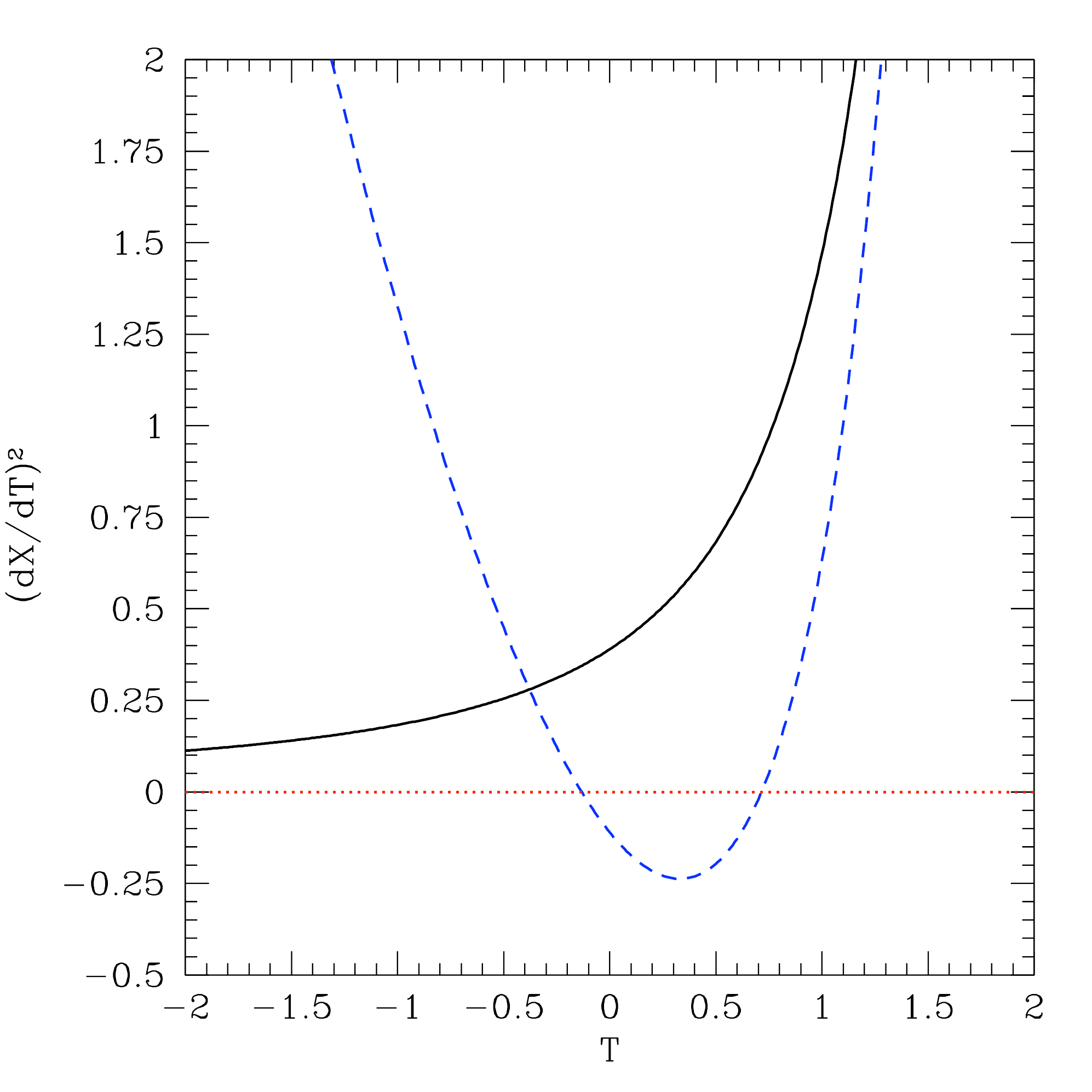}
\end{center}
\caption{The square of the dimensionless radial velocity $dX/dT$ as a
function of dimensionless time $T$ during the transition from
adiabatic inspiral to plunge.  The solid black curve shows the
left-hand side of Eq.~(\ref{E:dimKE}) for the numerical solution to
Eq.~(\ref{E:dimEOM}), while the dashed blue curve shows the right-hand
side of Eq.~(\ref{E:dimKE}) for this {\it same} numerical solution.
The two curves clearly differ, and the dashed blue curve becomes
negative when it falls below the dotted red line.  Since $(dX/dT)^2$
is positive definite, this indicates that the numerical solution
occupies a forbidden portion of $\{ X, T \}$ parameter space.}
\label{F:dimKE}
\end{figure}

We noted in Sec.~\ref{SS:geo} that the normalization of the 4-velocity
(\ref{E:4norm}) and the geodesic equation (\ref{E:GE}) provide
alternative equations of radial motion.  While they are equivalent for
geodesics, they differ once $\TE$ and $\TL$ become time dependent.
Ori and Thorne \cite{Ori:2000zn} solve the geodesic equation under the
assumption that $\TE$ and $\TL$ vary with proper time according to
Eq.~(\ref{E:xicon}).  What does this assumption imply for the norm of
the 4-velocity?  Expressed in the dimensionless variables of
Eq.~(\ref{E:dimvar}), Eq.~(\ref{E:1OT}) becomes
\begin{equation} \label{E:dimKE}
\left( \frac{dX}{dT} \right)^2 = -\frac{2}{3} X^3 - 2XT~.
\end{equation}
The left and right-hand sides of Eq.~(\ref{E:dimKE}) for the numerical
solution $X(T)$ of Eq.~(\ref{E:dimEOM}) are plotted in
Fig.~\ref{F:dimKE}.  The solid black curve, showing the left-hand side
$(dX/dT)^2$, is positive definite as one would expect.  The dashed
blue curve shows the right-hand side.  If Eq.~(\ref{E:dimKE}) was
satisfied, the two curves would be identical and the right-hand side
would be positive definite as well.  This is clearly not the case.

What does this mean physically?  According to Eq.~(\ref{E:xicon}), the
dimensionless variable $T$ is simultaneously proportional to the
proper time $\Ttau$, the energy $\chi$, and the angular momentum
$\xi$.  Values of $X(T)$ for which the right-hand side of
Eq.~(\ref{E:dimKE}) is negative correspond to values of $\{ \Tr, \TE,
\TL \}$ for which the effective potential $V(\Tr, \TE, \TL)$
exceeds the square of the energy $\TE^2$ in violation of
Eq.~(\ref{E:RV2}).  For example, at $T = 0$ the energy and angular
momentum equal their ISCO values, yet $X(0) > 0$ as seen in
Fig.~\ref{F:dimtraj}.  Since $V(\Tr_{\rm ISCO}, \TE_{\rm ISCO},
\TL_{\rm ISCO}) = \TE_{\rm ISCO}^2$ and the effective potential $V$ is
a monotonically increasing function of $X$ for $T = 0$, the right-hand
side of Eq.~(\ref{E:dimKE}) must be negative at $T = 0$ as seen in
Fig.~\ref{F:dimKE}.  The discrepancy between the left and right-hand
sides of Eq.~(\ref{E:dimKE}) is not a consequence of a breakdown in
the Taylor expansion of the effective potential for $R, \xi \gtrsim
1$.  Figure \ref{F:dimKE} shows that Eq.~(\ref{E:dimKE}) is violated
even for $X, T \simeq 1$, which corresponds to $R, \xi \ll 1$ when
$\eta \ll 1$.

\begin{figure}[t!]
\begin{center}
\includegraphics[width=3.5in]{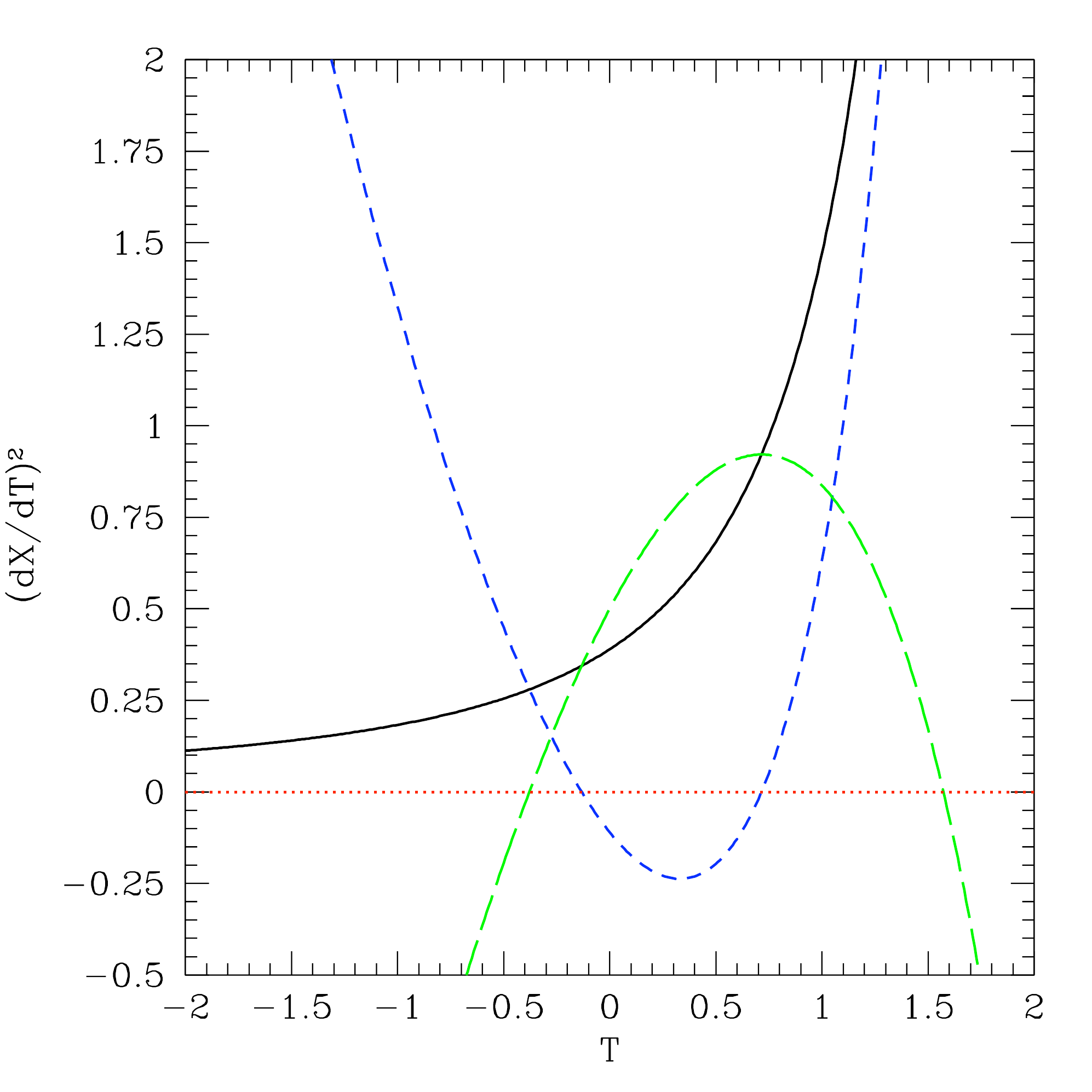}
\end{center}
\caption{The solid black, short-dashed blue, and dotted red curves
in this plot are identical to those in Fig.~\ref{F:dimKE}, but we have
also added the dimensionless difference between the energy and angular
momentum $Y(T)$ which is given by the long-dashed green curve.  The
short-dashed blue curve shows the first two terms on the right-hand
side of Eq.~(\ref{E:dimKEnew}), while the long-dashed green curve
shows our new third term.  Their sum equals the solid black curve,
implying that Eq.~(\ref{E:dimKEnew}) is now satisfied and the test
particle's 4-velocity is properly normalized.}\label{F:Y}
\end{figure}

How can we reconcile the normalization of the 4-velocity
(\ref{E:4norm}) and the geodesic equation (\ref{E:GE}) in the presence
of radiation reaction?  Although Eqs.~(\ref{E:circrad}) and
(\ref{E:xicon}) correctly specify the energy, angular momentum, and
their derivatives on a {\it circular} orbit at $\Tr_{\rm ISCO}$, there
is no reason to expect them to hold exactly when the radial velocity
is nonzero as it is during the transition.  If we relax the
requirement that $\chi = \xi$, we recover the third and fourth terms
on the right-hand side of Eq.~(\ref{E:1OT}) that vanished in the
previous treatment.  We can retain the dimensionless form of the
equations of motion by defining
\begin{equation} \label{E:Ydef}
\chi - \xi \equiv \eta^{6/5} (\chi - \xi)_0 Y~,
\end{equation}
where
\begin{equation} \label{E:Yscale}
(\chi - \xi)_0 = \alpha^{-4/5} (\beta\kappa)^{6/5} \left(
\frac{\partial V}{\partial \TL} \right)^{-1}~.
\end{equation}
The second-order equation (\ref{E:dimEOM}) remains unchanged to lowest
order in $\eta$, while Eq.~(\ref{E:dimKE}) gains an additional term to
become
\begin{equation} \label{E:dimKEnew}
\left( \frac{dX}{dT} \right)^2 = -\frac{2}{3} X^3 - 2XT + Y~.
\end{equation}
Instead of Eq.~(\ref{E:circrad}), which would imply that $Y$ is
constant throughout the transition, we evolve $Y$ according to
\begin{equation} \label{E:dYdT}
\frac{dY}{dT} = 2X~.
\end{equation}
This is precisely what is required to restore the consistency between
Eqs.~(\ref{E:dimEOM}) and (\ref{E:dimKEnew}) that exists for geodesic
motion.  Since Eq.~(\ref{E:dimEOM}) remains unchanged, its solution
$X(T)$ remains unchanged as well.

We can solve Eq.~(\ref{E:dYdT}) by inserting $X(T)$ into the
right-hand side and choosing the correct initial condition.  This
initial condition can be found by matching to the quasicircular
inspiral at early times.  During the inspiral, $\TE$ and $\TL$ are
given by
\cite{Bardeen:1972fi}
\begin{subequations} \label{E:ELgeo}
  \begin{eqnarray} \label{E:Egeo}
	\TE_c(\Tr) &=& \frac{1 - 2/\Tr + \Ta/\Tr^{3/2}}{\sqrt{1 -
	3/\Tr +2\Ta/\Tr^{3/2}}}~, \\ \label{E:Lgeo}
	\TL_c(\Tr) &=& \Tr^{1/2} \frac{1 - 2\Ta/\Tr^{3/2} +
	\Ta^2/\Tr^2}{\sqrt{1 - 3/\Tr +2\Ta/\Tr^{3/2}}}~.
  \end{eqnarray}
\end{subequations}
Taylor expanding about the ISCO,
\begin{eqnarray} \label{E:CXearly}
	\chi - \xi &\equiv& \TO_{\rm ISCO}^{-1}(\TE_c(r) - \TE_{\rm ISCO})
	- (\TL_c(r) - \TL_{\rm ISCO})
	\nonumber \\
	&\simeq& \frac{1}{6} \left( \TO^{-1} \frac{d^3\TE_c}{d\Tr^3} -
	\frac{d^3\TL_c}{d\Tr^3} \right)_{\rm ISCO} R^3~.
\end{eqnarray}
Using the definitions of $X$ and $Y$ in Eqs.~(\ref{E:dimR}) and
(\ref{E:Ydef}), this implies
\begin{eqnarray} \label{E:Yearly}
	Y &=& \frac{1}{6} \alpha^{-1} \left( \TO^{-1} \frac{d^3\TE_c}{d\Tr^3}
	- \frac{d^3\TL_c}{d\Tr^3} \right)_{\rm ISCO} \left(
	\frac{\partial V}{\partial \TL} \right)_{\rm ISCO} X^3
	\nonumber \\
	&=& -\frac{4}{3} X^3 \quad \quad T \to -\infty~.
\end{eqnarray}
Inserting Eqs.~(\ref{E:diminsp}) and (\ref{E:Yearly}) into the
right-hand side of Eq.~(\ref{E:dimKEnew}) shows that $dX/dT$ vanishes
as $T \to -\infty$ as required by Eq.~(\ref{E:diminsp}).  The
numerical solution $Y(T)$ with this initial condition is shown by the
long-dashed green curve in Fig.~\ref{F:Y}.  With the addition of this
new term to the right-hand side, Eq.~(\ref{E:dimKEnew}) is now
satisfied ensuring that the 4-velocity is normalized according to
Eq.~(\ref{E:masscon}).

Our new term $Y(T)$ is important for more than just mathematical
consistency.  Equation (\ref{E:Ydef}) suggests that the physical quantity
$\chi - \xi$ is proportional to $\eta^{6/5}$ during the transition,
making it higher order than self-force corrections that scale as
$\eta$.  However, as the test particle plunges into the horizon, the
asymptotic solution (\ref{E:dimplunge}) and Eq.~(\ref{E:dYdT}) imply
\begin{eqnarray} \label{E:YplungeT}
	Y(T) &=& Y(0) + \int_{0}^{T} \frac{dY}{dT^\prime} dT^\prime
	\nonumber \\
	&\simeq& Y(0) - \int_{0}^{T} \frac{12}{(\Tp - T^\prime)^2}
	dT^\prime \nonumber \\
	&\simeq& -\frac{12}{\Tp - T}
\end{eqnarray}
which diverges as $T \to \Tp$.  This divergence can be seen from the
limiting behavior of the three curves in Fig.~\ref{F:Y}.  Although the
solid black and short-dashed blue curves appear to converge as $T \to
\Tp$, their difference $Y(T)$ shown by the long-dashed green curve
in fact diverges in accordance with Eq.~(\ref{E:YplungeT}).  Does the
divergence of the dimensionless $Y(T)$ imply a similar divergence in
the physical quantity $\chi - \xi$ as the test particle plunges into
the black hole?

To answer this question, we must examine the validity of the
asymptotic solution (\ref{E:dimplunge}) as $T \to \Tp$.  Footnote 3 of
OT notes that the divergence of $X(T)$ results from a breakdown in the
dimensionless equation of motion (\ref{E:dimEOM}) when higher-order
terms in the Taylor expansion about the ISCO become important.  This
breakdown occurs at a coordinate radius $R_{\rm break}$, which
corresponds to a dimensionless radius $X_{\rm break} \propto
\eta^{-2/5}$ according to Eq.~(\ref{E:dimR}).  This in turn implies that 
$\Tp - T_{\rm break} \propto \eta^{1/5}$ by Eq.~(\ref{E:dimplunge})
and $Y_{\rm break} \propto \eta^{-1/5}$ from Eq.~(\ref{E:YplungeT}).
According to Eq.~(\ref{E:Ydef}), the test particle's energy during the
transition will receive a correction
\begin{equation} \label{E:DelEcorr}
\Delta \TE_{\rm norm} = \eta^{6/5} \TO_{\rm ISCO} (\chi - \xi)_0 Y~.
\end{equation}
When $Y \sim Y_{\rm break}$, this correction will be linearly
proportional to $\eta$ just like self-force corrections.  It will be
instructive to compare future self-force and time-domain perturbation
theory calculations with this analytic result.

\section{Maximal Spins} \label{S:maxspins}

Kerr black holes \cite{Kerr:1963ud} have spins $\Ta < 1$; objects with
larger spins are ``naked singularities'' unclothed by an event
horizon.  Penrose \cite{Penrose:1969pc} proposed that a cosmic censor
protects general relativity by preventing the formation of such naked
singularities, but this cosmic censorship conjecture has never been
proven in full generality.  Wald \cite{Wald:1974tp} determined that
maximally spinning black holes could not accrete {\it test} particles
that would drive them over the Kerr limit $\Ta = 1$, but Jacobson and
Sotiriou \cite{Jacobson:2009kt} recently showed that black holes with
spins $\delta \equiv 1 - \Ta \ll 1$ could be spun above this limit by
accreting particles with a {\it finite} energy $E = \eta M \TE$.
Barausse {\it et al.} \cite{Barausse:2010ka} showed that gravitational
radiation could not prevent some of these finite-mass-ratio mergers
from producing a naked singularity, but that self-force corrections
could be significant.  Given that finite-mass-ratio effects may
determine whether naked singularities can exist, it is worthwhile to
examine what happens to the transition region in the maximally
spinning limit.

The previous treatment of the transition from quasicircular inspiral to
plunge assumes that the effective potential $V$ and fluxes
$d\TE/d\Ttau$ and $d\TL/d\Ttau$ can be Taylor expanded in a
neighborhood of the ISCO.  However, as the black hole's spin
approaches the Kerr limit $a \to M$, $r_{\rm ISCO}$ and the outer
horizon $r_+$ both approach $M$ in Boyer-Lindquist coordinates.  This
suggests that it may not be possible to construct a neighborhood of
the ISCO that does not include the horizon.  To determine whether or
not such a neighborhood exists, we must compare $R_+ \equiv \Tr_{\rm
ISCO} - \Tr_+$ with $R_0$ in Eq.~(\ref{E:Rscal}) as $\delta \to 0$.
In this limit \cite{Bardeen:1972fi},
\begin{subequations} \label{E:Rlim}
  \begin{eqnarray} \label{E:ISCOlim}
	\Tr_{\rm ISCO} &\to& 1 + (4\delta)^{1/3}~, \\ \label{E:EHlim}
	\Tr_+ &\to& 1 + (2\delta)^{1/2}~,
  \end{eqnarray}
\end{subequations}
implying $R_+ \propto \delta^{1/3}$.

\begin{figure}[t!]
\begin{center}
\includegraphics[width=3.5in]{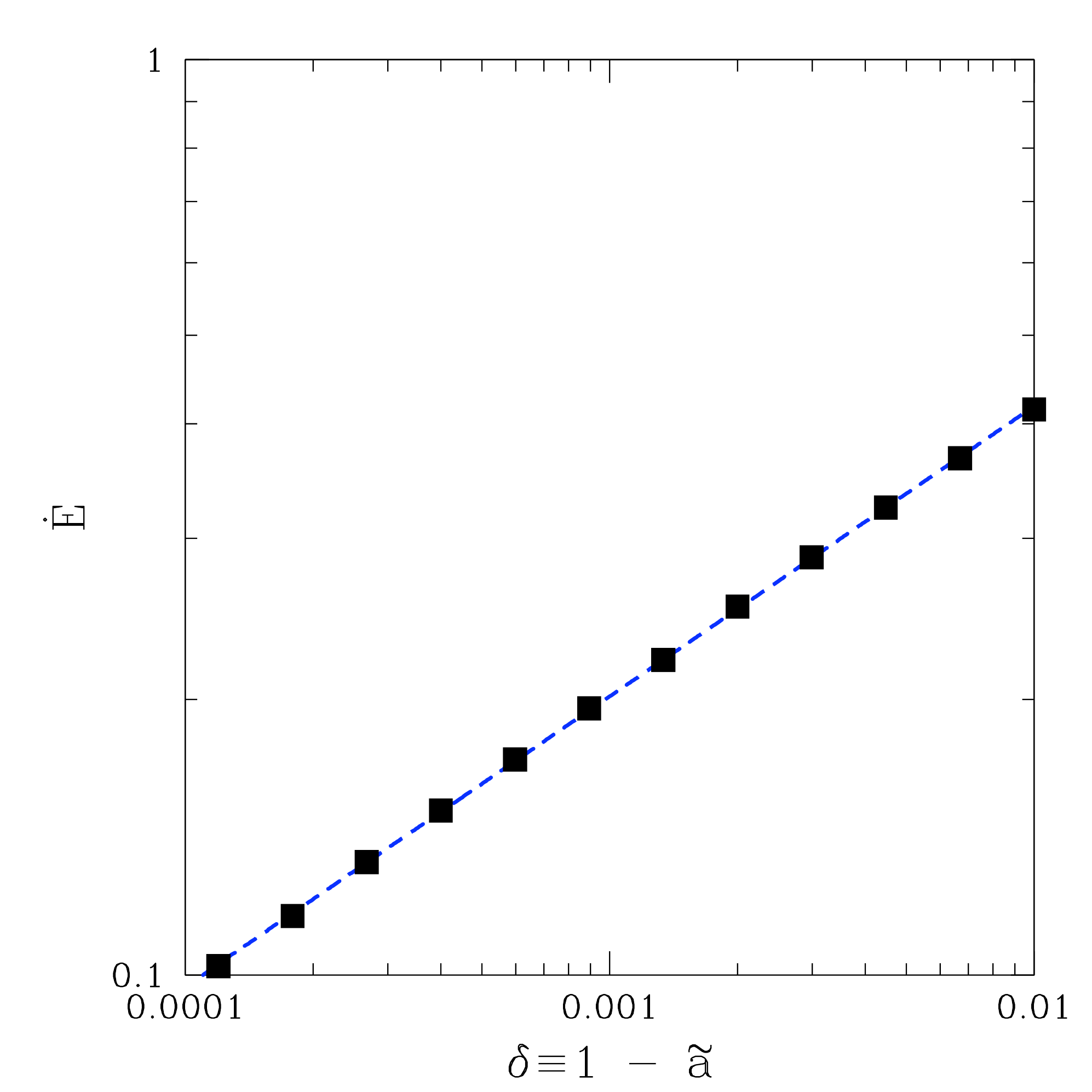}
\end{center}
\caption{The general-relativistic correction $\RC$ to the Newtonian
quadrupole-moment formula for the GW luminosity as a function of
black-hole spin $\delta \equiv 1 - \Ta$.  The points have been
calculated using the black-hole perturbation theory code GREMLIN
\cite{Hughes:1999bq} and fitted with a power law $\RC = A \delta^m$.}
\label{F:RC}
\end{figure}

The behavior of $R_0$ as $\delta \to 0$ depends on the behavior of the
fluxes $d\TE/d\Ttau$ and $d\TL/d\Ttau$.  These in turn depend on
$\RC$, defined by
\begin{equation} \label{E:RC}
\dot{E}_{\rm GW} = \frac{32}{5} \eta^2 \TO^{10/3} \RC
\end{equation}
as the general-relativistic correction to the Newtonian
quadrupole-moment formula for the GW luminosity
\cite{Finn:2000sy}.  The ISCO values of this correction $\RC$ were
calculated for spins $-0.99 \leq \Ta \leq 0.999$ in
\cite{Finn:2000sy}, and have been calculated down to $\delta = 10^{-4}$
\cite{HughesComm} using the GREMLIN (Gravitational
Radiation in the Extreme Mass Ratio Limit) code presented in
\cite{Hughes:1999bq}.  We have fitted these calculated values to a power law
$\RC = A \delta^m$ as shown in Fig.~\ref{F:RC}.  The best-fit
parameters for this power law are $A = 1.80$, $m = 0.317$.  The
summation over spheroidal harmonics of the Weyl scalar $\psi_4$ needed
to compute $\RC$ converges very slowly in the limit $\delta \to 0$, so
our best-fit parameters should be regarded with caution until they can
be confirmed by a technique better suited to this limit.  Chrzanowski
\cite{Chrzanowski:1976jy} estimated that
\begin{equation} \label{E:Chrz}
\dot{E}_{\rm GW} \sim \eta^2 R_+~,
\end{equation}
which also suggests that $m \sim 1/3$.  The rough agreement between
our numerical fit and Chrzanowski's estimate gives us some confidence
that $m \simeq 1/3$ is close to the correct value.

The scale $R_0$ of the transition region depends on the relativistic
correction $\RC$ through $\kappa$, which was defined in
Eq.~(\ref{E:kappadef}) and can be expressed as
\begin{equation} \label{E:kappa2}
\kappa = \frac{32}{5} \left( \TO^{7/3} \frac{d\Tt}{d\Ttau}
\RC \right)_{\rm ISCO}~.
\end{equation}
For circular equatorial Kerr geodesics, as $\delta \to 0$,
\begin{subequations} \label{E:max1}
  \begin{eqnarray} \label{E:Omax}
	\TO &\to& \frac{1}{2} \left[ 1 - \frac{3}{4} (4\delta)^{1/3} \right]~,
	\\ \label{E:dtdtaumax}
	\frac{d\Tt}{d\Ttau} &\to& \frac{4}{\sqrt{3}} (4\delta)^{-1/3}~,
	\\ \label{E:kappamax}
	\kappa &\to& \frac{16}{5\sqrt{3}} A \delta^{m - 1/3}~,
	\\ \label{E:alphamax}
	\alpha &\to& 1~, \\ \label{E:betamax}
	\beta &\to& \frac{\sqrt{3}}{2} (4\delta)^{1/3}~, \\ \label{E:dVdLmax}
	\frac{\partial V}{ \partial \TL} &\to& 
	\frac{4}{\sqrt{3}} (4\delta)^{1/3}~,
  \end{eqnarray}
\end{subequations}
implying
\begin{subequations} \label{E:max2}
  \begin{eqnarray} \label{E:R0max}
	R_0 &\propto& \delta^{2m/5}~, \\ \label{E:tau0max}
	\tau_0 &\propto& \delta^{-m/5}~, \\ \label{E:chi0max}
	(\chi - \xi)_0 &\propto& \delta^{6m/5 - 1/3}~,
  \end{eqnarray}
\end{subequations}
from the definitions in Eqs.~(\ref{E:scal}) and (\ref{E:Yscale}).

\begin{figure}[t!]
\begin{center}
\includegraphics[width=3.5in]{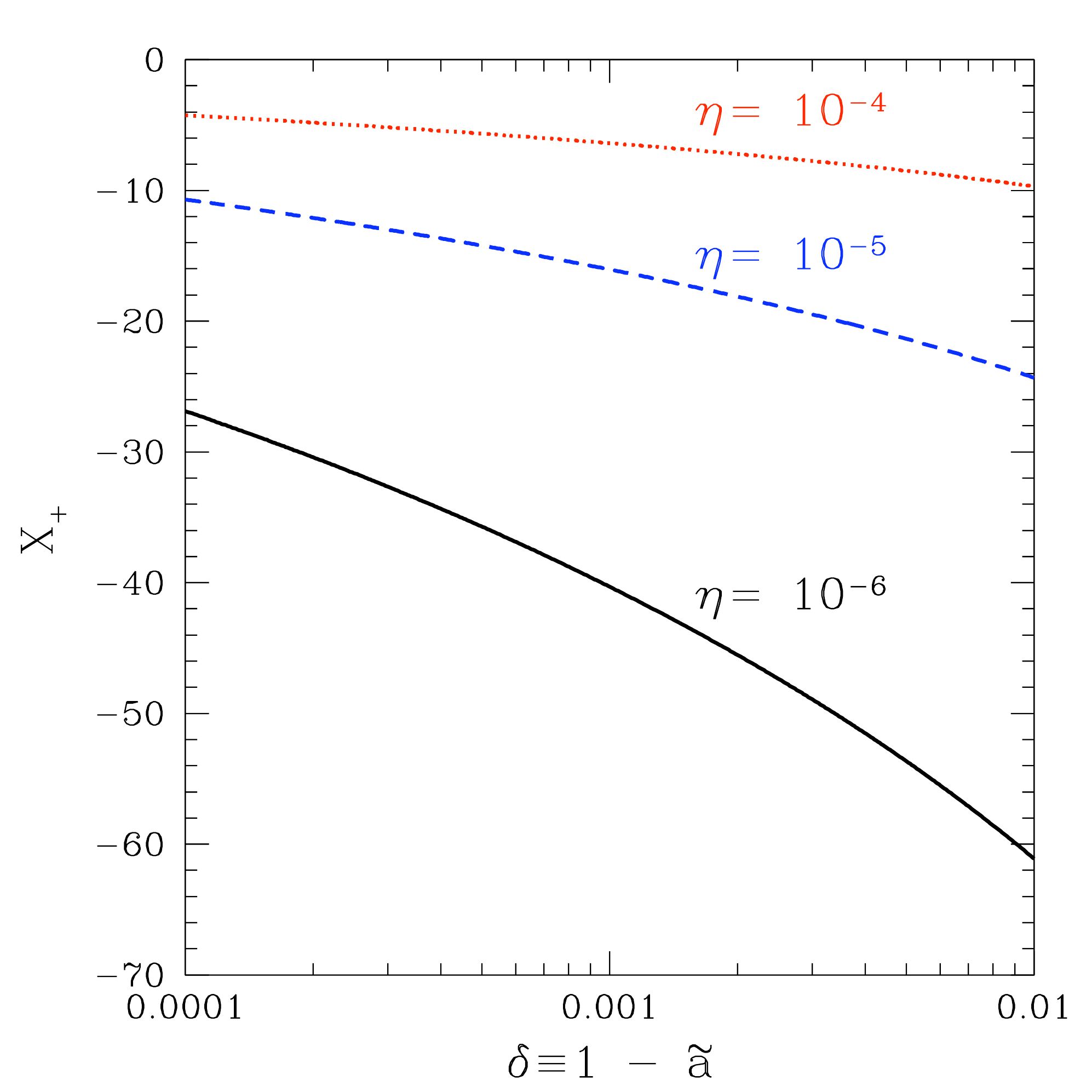}
\end{center}
\caption{The dimensionless horizon radius $X_+ = -\eta^{-2/5}R_+/R_0$
as a function of $\delta \equiv 1 - \Ta$.  The dotted red, dashed
blue, and solid black curves correspond to mass ratios $\eta =
10^{-4}$, $10^{-5}$, and $10^{-6}$ respectively.  $X_+ \to 0$ as
$\delta \to 0$ because the relativistic correction $\RC$ to the energy
radiated in GWs has a power-law index $m < 5/6$.}
\label{F:EH}
\end{figure}

According to Eq.~(\ref{E:dimR}), the transition region will be cut off
by the horizon at a dimensionless radius
\begin{equation} \label{E:Xcut}
X_+ = -\eta^{-2/5} \frac{R_+}{R_0} \propto -\eta^{-2/5} \delta^{1/3 - 2m/5}
\end{equation}
as $\delta \to 0$.  We plot $X_+(\delta)$ for several mass ratios and
our best-fit value of $m$ in Fig.~\ref{F:EH}.  Since $m < 5/6$, $X_+
\to 0$ in the limit of large spins.  One should not expect GW emission
beyond $X_+$, so Eq.~(\ref{E:final}) overestimates the additional
energy and angular momentum radiated during the transition.  In the
limit of large spins, $\Delta \TE_{\rm tr}$ and $\Delta \TL_{\rm tr}$
will be suppressed by a factor $T_0/\Tp \simeq 0.21$, where $X(T_0) = 0$.

Despite this suppression, the extra energy and angular momentum
radiated during the transition remain proportional to $\kappa \tau_0
\propto \delta^{4m/5 - 1/3}$, which diverges for $m < 5/12$.  Our
numerical fit shown in Fig.~\ref{F:RC} suggests that $m$ satisfies
this inequality.  However, this divergence might not be physical; it
could merely reflect a breakdown of the Taylor expansion of the
equation of motion given in Eq.~(\ref{E:2OT}).  The first two terms on
the right-hand side of this equation are proportional to $\eta^{4/5}$,
while Eq.~(\ref{E:Ydef}) shows that the third term is proportional to
$\eta^{6/5}$.  This higher-order dependence on $\eta$ justifies
neglecting this third term for modest spins, but we must also consider
how each term depends on $\delta$ in the maximally spinning limit.  In
this limit, the first two terms are proportional to $\delta^{4m/5}$
according to Eqs.~(\ref{E:max1}) and (\ref{E:max2}) while the third
term is proportional to $\delta^{6m/5 - 1/3}$.  For $m < 5/6$, this
third term will dominate over the first two terms, indicating a
breakdown of the original dimensionless equation of motion (\ref{E:dimEOM}).
For completeness, we note that we have neglected an additional term
\begin{equation} \label{E:4term}
-\frac{1}{12} \left( \frac{\partial^4 V}{\partial \Tr^4} \right)_{\rm ISCO} R^3
\end{equation}
at $\mathcal{O}(\eta^{6/5})$ on the right-hand side Eq.~(\ref{E:2OT}).
This term is proportional to $\delta^{6m/5}$ as $\delta \to 0$ and
thus can always be neglected.

\begin{figure}[t!]
\begin{center}
\includegraphics[width=3.5in]{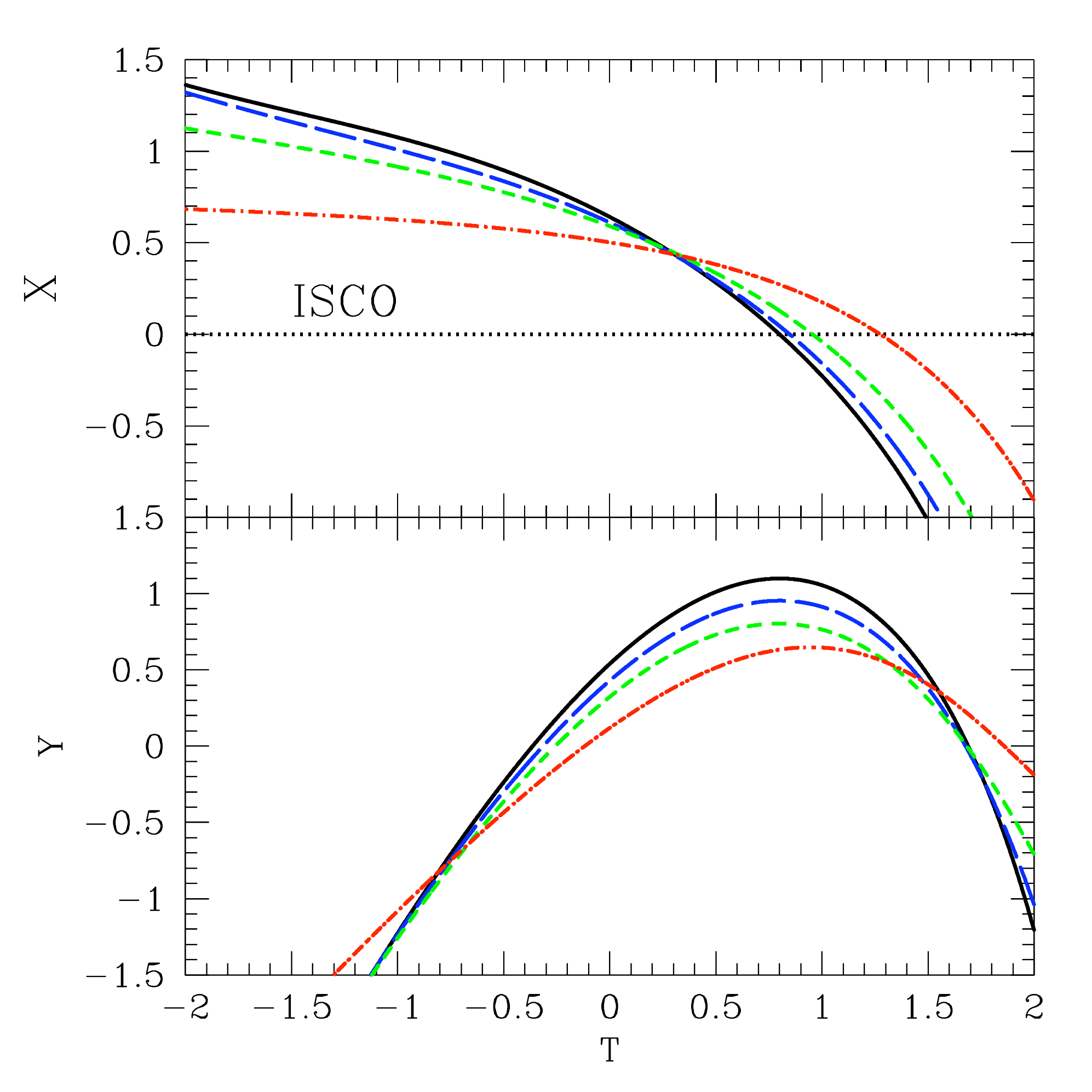}
\end{center}
\caption{{\it Upper panel:} The dimensionless radius $X$ as a function
of dimensionless time $T$ in the limit $\delta \equiv 1 - \Ta \to
0$. The solid black curve is the OT solution, while the long-dashed
blue, short-dashed green, and dot-dashed red curves show our solutions
with $\eta = 10^{-3}$ and $\delta = 10^{-2}$, $10^{-4}$ and $10^{-6}$
respectively.  {\it Lower panel:} The dimensionless difference $Y$
between the energy and angular momentum for the solutions shown in the
upper panel.}
\label{F:3T}
\end{figure}

Including the third term in Eq.~(\ref{E:2OT}) on the right-hand side
of Eq.~(\ref{E:dimEOM}) yields
\begin{equation} \label{E:dim3T}
\frac{d^2X}{dT^2} = - X^2 - T + \eta^{2/5} CY~,
\end{equation}
where
\begin{equation} \label{E:3TC}
C \equiv -\frac{1}{2} \alpha^{-3/5} (\beta \kappa)^{2/5} \TO
\frac{\partial^2 V}{\partial \TE \partial \Tr} \left(
\frac{\partial V}{\partial \TL} \right)^{-1} \propto \delta^{2m/5 - 1/3}
\end{equation}
diverges as $\delta \to 0$ for $m < 5/6$.  $Y$ must evolve according
to
\begin{equation} \label{E:Y3T}
\frac{dY}{dT} = 2X + 2\eta^{2/5} CY \frac{dX}{dT}
\end{equation}
to preserve the normalization the 4-velocity given by
Eq.~(\ref{E:dimKEnew}).  The product $\eta^{2/5}C$ thus measures the
deviation of $X(T)$ from the OT solution in the limit of large spins.
We plot our new solutions $X(T,\eta^{2/5}C)$ for a mass ratio $\eta =
10^{-3}$ and spins $\delta = 10^{-2}$, $10^{-4}$ and $10^{-6}$ in
Fig.~\ref{F:3T}.  Although our solutions diverge from the OT solution
as $\eta^{2/5}C$ increases, the test particle crosses the ISCO $X = 0$
(shown by the horizontal dotted line) at a {\it later} dimensionless
time $T$.  This implies that more energy and angular momentum are
radiated and the divergence in $\Delta \TE_{\rm tr}$ and $\Delta
\TL_{\rm tr}$ as $\delta \to 0$ cannot be avoided.  One must look to
finite-mass ratio effects beyond the scope of this paper to eliminate
this unphysical result.

One such effect that might provide a solution to this problem is the
spinning down of the black hole due to the extraction of angular
momentum by the superradiant scattering of GWs earlier in the inspiral
\cite{Kesden:2009ds}.  This effect implies that even an initially
maximally spinning black hole will have $\delta \propto \eta$ by the
time the test particle reaches the transition region.  The additional
energy and angular momentum $\Delta \TE_{\rm tr}, \TL_{\rm tr} \propto
\eta^{4/5} \delta^{4m/5 - 1/3} \propto \eta^{4m/5 + 7/15}$ would then
remain finite even if $\delta = 0$ initially.

How does the energy and angular momentum radiated during the
transition affect the spin of the final black hole produced in the
merger?  If we assume that energy and angular momentum are conserved
after the end of the transition, the final spin will be
\begin{eqnarray} \label{E:af}
	\Ta_f &=& \frac{\Ta + \eta(\TL_{\rm ISCO} - \Delta \TL_{\rm tr})}
	{[1 + \eta(\TE_{\rm ISCO} - \Delta \TE_{\rm tr} +
	\Delta \TE_{\rm norm})]^2}
	\nonumber \\
	&\simeq& \Ta + \eta (\Delta \Ta_{\rm ISCO} + \Delta \Ta_{\rm tr}
	+ \Delta \Ta_{\rm norm})
\end{eqnarray}
for $\eta \ll 1$, where in the limit $\delta \to 0$
\begin{subequations} \label{E:max3}
  \begin{eqnarray} \label{E:ISCOmax}
	\Delta \Ta_{\rm ISCO} &\equiv& \TL_{\rm ISCO} - 2\TE_{\rm ISCO}
	\to \frac{\sqrt{3}}{2} (4\delta)^{2/3},
	\\ \label{E:Trmax}
	\Delta \Ta_{\rm tr} &\equiv& -(1 - 2\TO_{\rm ISCO})\Delta \TL_{\rm tr}
	\propto \eta^{4/5} \delta^{4m/5},
	\\ \label{E:Normmax}
	\Delta \Ta_{\rm norm} &\equiv& -2\Delta \TE_{\rm norm}
	\propto \eta^{6/5} \delta^{6m/5 - 1/3}.
  \end{eqnarray}
\end{subequations}
Although our new correction $\Delta \Ta_{\rm norm}$ to the black
hole's final spin is subdominant in $\eta$, for $m < 5/6$ it becomes
the dominant correction as $\delta \to 0$.  Since $\Delta \TE_{\rm
norm} > 0$ when the particle crosses the horizon in the limit $\delta
\to 0$, our analysis suggests that gravitational radiation in the
transition region will {\it not} promote the formation of naked
singularities.  This supports our earlier result in
\cite{Kesden:2009ds}, where we showed that the superradiant scattering
of GWs emitted at $\Tr > \Tr_{\rm ISCO}$ would also reduce the final
spin for $\Ta \gtrsim 0.998$.

\section{Discussion} \label{S:disc}

The existence of an ISCO is one of the most distinctive features of a
black hole.  Although the presence of an ISCO indirectly affects the
luminosity and spectra of accreting black holes, it is more cleanly
probed by the GWs emitted as test particles (compact objects like
white dwarfs, neutron stars, and stellar-mass black holes) plunge into
supermassive black holes.  Such GWs are a primary source for the
proposed space-based GW detector LISA.  The LISA detection strategy
relies on convolving observations with theoretically determined GW
templates.  Understanding how a test particle's position, energy, and
angular momentum evolve near the ISCO is an important first step
toward constructing these templates.

Previous studies \cite{Ori:2000zn,Buonanno:2000ef} identified a
transition region near the ISCO where neither the quasicircular
approximation nor the assumption of geodesic motion are valid.  They
solved the radial equation of motion in this region by Taylor
expanding it about the ISCO.  Ori and Thorne (2000) \cite{Ori:2000zn}
fixed the energy and angular momentum fluxes to their ISCO values,
while Buonanno and Damour (2000) \cite{Buonanno:2000ef} worked with
the Hamilton equations of motion and did not need to specify an
explicit energy flux.  The dimensionless solution $X(T)$ obtained by
both groups can be rescaled to determine the evolution of the test
particle's position, energy, and angular momentum for arbitrary mass
ratios and spins.  The simplicity of their approach and the
universality of their solution are highly appealing, but a closer
examination reveals that their solution does not properly normalize
the particle's 4-momentum.

We undertook this study to see whether this technical problem could be
easily remedied, or was a symptom of a more serious flaw in their
approach.  We found that by introducing a correction $\Delta \TE_{\rm
norm}$ to the particle's energy at a higher order in the mass ratio
$\eta \ll 1$, we could properly normalize the 4-momentum without
altering their universal solution $X(T)$ at lowest order.  This
relatively modest correction increases our confidence in this particle
trajectory, and validates its use as a source for the construction of
GW waveforms as in \cite{Sundararajan:2010sr}.

In addition to its role in constructing GW templates, gravitational
radiation during the transition also affects whether a test-particle
merger can produce a naked singularity by increasing a black hole's
spin above the Kerr limit $\Ta = 1$.  Our calculation of the energy
and angular momentum flux at the ISCO as $\delta \equiv 1 - \Ta \to 0$
suggests that the total energy $\Delta \TE_{\rm tr}$ and angular
momentum $\Delta \TL_{\rm tr}$ radiated during the transition diverge
in this limit.  This divergence cannot be physical, and must therefore
be moderated by high-spin corrections beyond the scope of this paper.
Despite this divergence, the change $\Delta \Ta_{\rm tr}$ in the black
hole's spin due to this radiation remains finite in magnitude and
negative in sign.  It therefore reduces the likelihood that a naked
singularity will be produced, as does our new correction $\Delta
\Ta_{\rm norm}$ which is also negative and becomes dominant as $\delta
\to 0$.

Throughout this paper, we have neglected self-force corrections to the
particle's energy even though they are formally lower order in $\eta$
than our correction $\Delta \TE_{\rm norm} \propto \eta^{6/5}$.  Far
from the ISCO, where $|Y| \gg 1$, our correction should dominate
self-force effects despite its scaling with $\eta$.  Near the ISCO
however, self-force corrections should be significant.  Reliable
calculations of the self-force do not yet exist near the ISCOs of
highly spinning black holes.  Comparing such self-force calculations
with our analysis as they become available is a subject for future
work.  Another important future test of our analysis would be to use
time-domain perturbation theory to calculate the energy and angular
momentum radiated by a particle whose position is given by an
appropriate rescaling of the solution $X(T)$.  Agreement with our
predicted solution $Y(T)$ would strongly support our conjecture that
proper normalization of the 4-momentum can be used to predict GW
emission.  Comparison with numerical relativity may soon be possible
as well, since simulations with mass ratios as small as $\eta = 0.01$
have recently been performed \cite{Lousto:2010ut}.  Our predictions
could be even further improved by including the gravitational
radiation emitted near the horizon \cite{Mino:2008at}.  We look
forward to comparing our calculation to these alternative theoretical
approaches in the near future, and comparing all these predictions to
GW observations in the hopefully not too distant future.

\vspace{0.3cm}

{\bf Acknowledgements.} I would like to thank Sterl Phinney for his
advice during the initial stages of this project, and Scott Hughes for
insight into the emission of gravitational radiation in the limit of
large spins.  Jeandrew Brink, Marc Favata, Chris Hirata, Guglielmo
Lockhart, Samaya Nissanke, and David Tsang also offered helpful
comments on this work.

\end{document}